\documentclass[prl,aps,twocolumn,showpacs,amssym,floatfix]{revtex4} 
\usepackage{graphicx,psfrag,amsmath,amssymb,float}
\usepackage{epstopdf}
\usepackage{color}
\input{epsf}
\usepackage{graphicx,natbib}
\newcommand{\be}{\begin{equation}}
\newcommand{\ee}{\end{equation}}
\newcommand{\bea}{\begin{eqnarray}}
\newcommand{\eea}{\end{eqnarray}}

\begin{document}
\title{Quantum and thermal transitions out of the supersolid phase of a 2D quantum antiferromagnet}
\author{Nicolas Laflorencie}
\affiliation{Insitute of Theoretical Physics, \'Ecole Polytechnique F\'ed\'erale de Lausanne, Switzerland.}
\author{Fr\'ed\'eric Mila}
\affiliation{Insitute of Theoretical Physics, \'Ecole Polytechnique F\'ed\'erale de Lausanne, Switzerland.}
\begin{abstract}
We investigate the thermodynamic properties of a field-induced supersolid phase in a 2D quantum antiferromagnet model. 
Using quantum Monte Carlo simulations, a very rich phase diagram is mapped out in the temperature - magnetic field plane, with an extended supersolid region where a diagonal (solid) order coexists with a finite XY spin stiffness (superfluid). The various quantum and thermal transitions out of the supersolid state are characterized.
Experimental consequences in the context of field-induced magnetization plateau materials are briefly discussed.
\end{abstract}
\maketitle
\noindent
{\it{Introduction---}}
Since the early proposals of a possible supersolid (SS) state of matter~\cite{Penrose56-Andreev69}, where both diagonal solid and off-diagonal superfluid (SF) long range orders coexist, the existence of this new exotic state of matter at low temperature has been intensively discussed in $^4$He~\cite{Kim04,Balibar06,Review,Pollet07}.
For lattice models, bosons have revealed an interesting tendency towards the formation of a stable (lattice) SS phase on frustrated networks like for instance the triangular lattice~\cite{Triangular,Prokof}.
Perhaps a more promissing route for the search of such an exotic state of matter points towards gapped quantum magnets in an external magnetic field where, using a hard-core boson representation for triplet rungs~\cite{frederic98-giam99}, a diagonal order is equivalent to a Bose solid which breaks the lattice translational symmetry and an off-diagonal order represents a finite fraction of condensed bosons which breaks the U(1) symmetry. 

Several Mott insulators are known to exhibit field induced Bose-Einstein condensation of the triplets~\cite{hanpurple,TiOCl}, as well as magnetization plateaus where an incompressible commensurate crystal of triplets is formed~\cite{Oshikawa97,Borate,Rice02}. 
However, a direct continuous (second order) transition between the U(1) symmetry breaking condensate and the charge density ordered plateau state would violate the Ginsburg-Landau paradigm. Therefore, one expects either a first order transition, or an exotic deconfined critical point~\cite{Senthil}, or an intermediate region where both competing orders might coexist, namely a SS state. Thus, it is natural to consider quantum magnets as serious candidates to achieve a SS state in the vicinity of a magnetization plateau. And indeed a SS phase has been recently reported for a spin-$\frac{1}{2}$ bilayer system~\cite{NgLee06}.

So far, most of the theoretical investigations have focused on the existence and the stability of the SS at $T=0$, leaving unanswered the question of the nature of the quantum and thermal transitions out of the SS. Note however that the thermal melting of the SS via two distinct transitions (Kozterlitz-Thouless and 3-state Potts) on the triangular lattice was reported in Ref.~\onlinecite{Prokof}.

\begin{figure}[!ht]
\begin{center}
\includegraphics[width=0.85\columnwidth,clip]{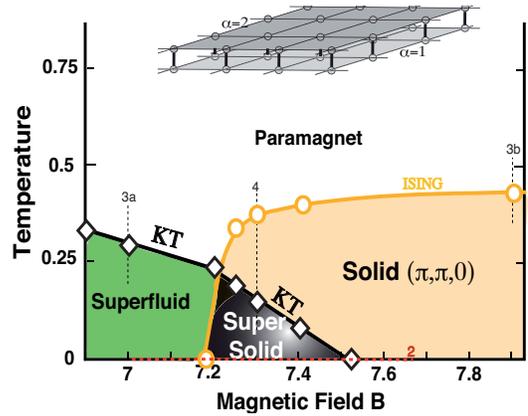}
\end{center}
\caption{(color online) Magnetic field - temperature phase diagram of the bilayer system Eq.~(\ref{eq:model}) depicted on the top. The supersolid region (dark grey) is displayed together with its neighboring phases: superfluid, solid, paramagnet. The critical lines belong to different universality classes, as indicated on the plot. The transition points (various symbols) result from quantum Monte Carlo simulations and finite size scaling analysis (see text and other plots). The thin dashed lines are constant $B$ or $T$ scans discussed below.}
\label{PhDgT}
\end{figure}
%

In this Letter, using the stochastic series expansion (SSE) quantum Monte Carlo algorithm~\cite{Sandvik02}, we explore the properties of the SS state of the quantum magnet model introduced in Ref.~\onlinecite{NgLee06}, with emphasis on a precise characterization of the quantum and thermal phase transitions out of the SS state.
The resulting magnetic field - temperature phase diagram is shown in Fig.~\ref{PhDgT}. 
As we shall see below, both $T=0$ quantum phase transitions (SF-SS and SS-solid) as well as the thermal transition (SS-paramagnet) display special features of direct experimental relevance.
In particular, the melting of the SS is a two-step process in contrast to standard transitions usually observed in quantum antiferromagnets.

\begin{figure}[!ht]
\begin{center}
\includegraphics[width=0.9\columnwidth,clip]{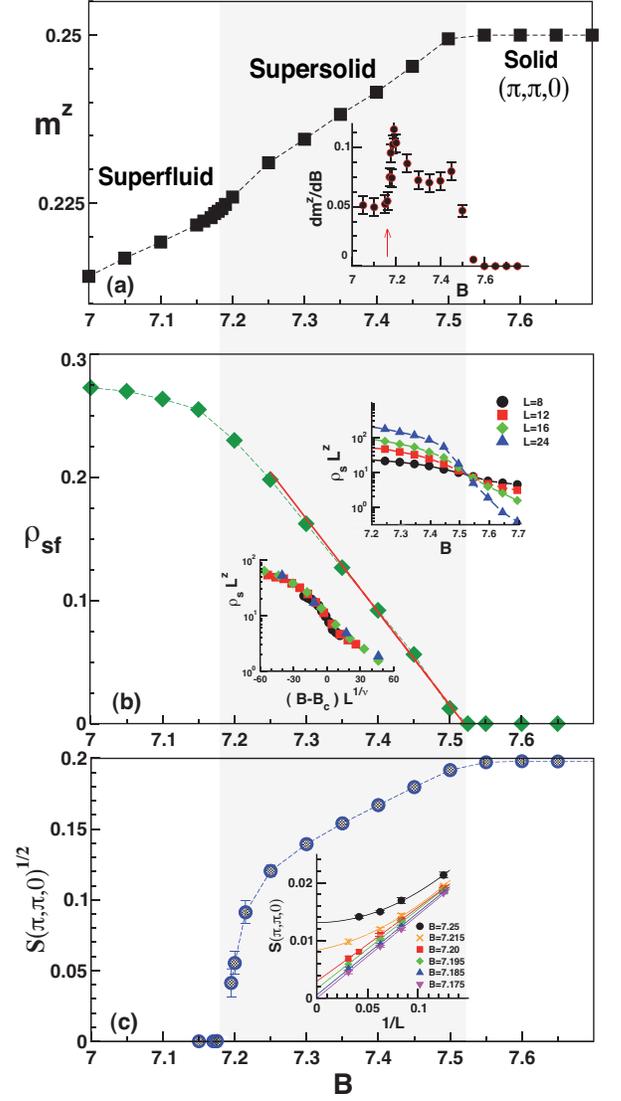}
\end{center}
\caption{Ground state properties of the model (\ref{eq:model}) focused on the superfluid, supersolid (light grey region) and solid regimes [scan 2]. The three main panels show infinite size extrapolations of quantum Monte Carlo estimates versus the external field $B$, obtained for $T=(2L)^{-1}$ with $8\le L\le 32$. (a) The uniform magnetization $m^{z}$ grows up to a plateau $m^{z}=0.25$. Inset (a): spin susceptibility $\chi(B)={\rm{d}}m^{z}/{\rm{d}}B$. (b) The spin stiffness $\rho_{\rm sf} $ decays in the supersolid regime and vanishes linearly at $B_{\rm s} =7.52$ like $\rho_{\rm sf} \simeq 0.757(B-B_{\rm s} )$ (straight line). Upper inset (b): Universal crossing of $\rho_{s}L^{z}$ at the critical field $B_{\rm s} $ with $z=2$. Lower inset (b): Data collapse of $\rho_{s}L^{z}$ versus $(B-B_{\rm s} )L^{1/\nu}$ obtained with $z=2$ and $\nu=1/2$. (c) The solid order parameter $S(\pi,\pi,0)^{1/2}$ grows up for $B> 7.175$ and saturate in the solid phase. Inset (c) Finite size scaling of $S(\pi,\pi,0)$ plotted versus $1/L$ for various fields $B$ indicated on the plot; the lines are quadratic fits in $L^{-1}$.}
\label{T0}
\end{figure}

\noindent
{\it{Zero temperature properties---}}
We consider the following bilayer spin-$\frac{1}{2}$ XXZ Hamiltonian
\begin{eqnarray}
{\cal{H}}=J\sum_{\langle i j\rangle,\alpha=1,2}\left(S_{i,\alpha}^{x}S_{j,\alpha}^{x}+S_{i,\alpha}^{y}S_{j,\alpha}^{y}+
\Delta S_{i,\alpha}^{z}S_{j,\alpha}^{z}\right)\nonumber \\
+J_{\perp}\sum_{i}{\bf{S}}_{i,1}\cdot{\bf{S}}_{i,2} -g\mu_B B\sum_{i,\alpha=1,2}S_{i,\alpha}^{z},
\label{eq:model}
\end{eqnarray}
where we set $g\mu_B=1$ and choose the following parameters $J=1$, $\Delta=3.3$ and $J_{\perp}=3.45$, for which an extended zero temperature SS state was found in Ref.~\cite{NgLee06}. 
Before discussing the finite temperature melting of the SS, we first characterize the ground state (GS) properties, using large scale SSE simulations on finite systems ($L\times L\times 2$) at low enough temperature: we have checked that using $T= (2L)^{-1}$ insures that we focus on the GS properties.
A diagonal order at a vector ${\bf q}=(q_x,q_y,q_z)$ will be signaled by the saturation to a finite value of the static structure factor per site
$
S({\bf q})={\langle{\left(\sum_{\bf r} \exp(i{\bf q} \cdot{\bf r})S_{\bf r}^{z}\right)^2}\rangle}/N^{2},
$
where the sum over ${\bf r}$ runs over $r_x=1,\hdots,L$; $r_y=1,\hdots,L$; $r_z=1,2$ and $N=2L^2$ is the total number of sites on the bilayer.
For the superfluidity, in contrast to Ref.~\onlinecite{NgLee06} where the condensate density was calculated,
here we concentrate on the spin stiffness $\rho_{\rm sf} $ of the U(1) off-diagonal ordering in the $xy$ plane since
this is the order parameter at finite temperature. It is measured via the fluctuations of the winding number~\cite{Ceperley-Kawashima} in the SSE simulation.
Using a hard-core boson mapping for the triplets~\cite{frederic98-giam99}, a finite value for $S({\bf q})$ corresponds to a commensurate charge density wave state (incompressible) for the bosons, and a non zero spin stiffness $\rho_{\rm sf}$ must be regarded as a finite SF density,
corresponding to a gapless (compressible) SF state. The total density of effective bosons is measured via the average magnetization per site $m^{z}=\langle\sum_{\bf r}S_{r}^{z}\rangle/N$, shown in Fig.~\ref{T0} (a) after an infinite size extrapolation with $L=8,12,16,24,32$. Upon increasing the field, $m^{z}$ grows coutinuously up to a half saturated magnetization plateau at  $m^{z}=0.25$. This gapped plateau state, called solid $(\pi,\pi,0)$, displays diagonal order at ${\bf q}=(\pi,\pi,0)$, as shown in Fig.~\ref{T0} (c).
When decreasing the field from the plateau, there is a gapless SF phase with a finite stiffness $\rho_{\rm sf}$ (see Fig.~\ref{T0} (b)).
Interestingly the solid $(\pi,\pi,0)$ is also present in this SF regime (see Fig.~\ref{T0} (c)), thus achieving a SS phase over a broad regime of magnetic field with both $\rho_{s}\neq 0$ {\it{and}} $S(\pi,\pi,0)\neq 0$. At lower fields, the solid is not stable anymore and a standard SF state with only $\rho_{s}\neq 0$ is recovered.
%

The transition between the SF and the SS can already be observed by looking at the field-induced magnetization curve in Fig.~\ref{T0} (a). Indeed, it displays a clear anomaly at the onset of the transition, as evidenced by a feature in the spin susceptibility $\chi(b)={\rm{d}}m^{z}/{\rm{d}}B$ (pointed by an arrow in the inset of Fig.~\ref{T0} (a)). 
This quantum phase transition SF-SS, driven by the external field $B$, occurs for $B_{\rm ss} =7.175(5)$. Its universality class seems to be consistent with 
3D (2+1) Ising as the critical decay of the structure factor $S(\pi,\pi,0)\sim L^{-2\beta/\nu}$ at $B_{\rm ss} $ (shown as a straight line vs $L^{-1}$ in the inset of 
Fig.~\ref{T0} (c)) is compatible with the 3D Ising value $2\beta/\nu=1.037$~\cite{Blote95}. 
The transition between the SS and the solid occurs when the stiffness vanishes, at $B_{\rm s} =7.52(1)$, as shown in Fig.~\ref{T0} (b). We find an excellent agreement with the SF-insulator universality class~\cite{Fisher89-Alet04} for this transition since the finite size scaling analysis is fully combatible with the mean-field exponents $z=2, \nu=1/2, \beta=1/2$, as demonstrated in the inset of Fig.~\ref{T0} (b).

\noindent
{\it{Finite temperature properties---}}
We now turn to the finite temperature properties of the SS and discuss in more details the phase diagram presented in Fig.~\ref{PhDgT}.  
\begin{figure}[!ht]
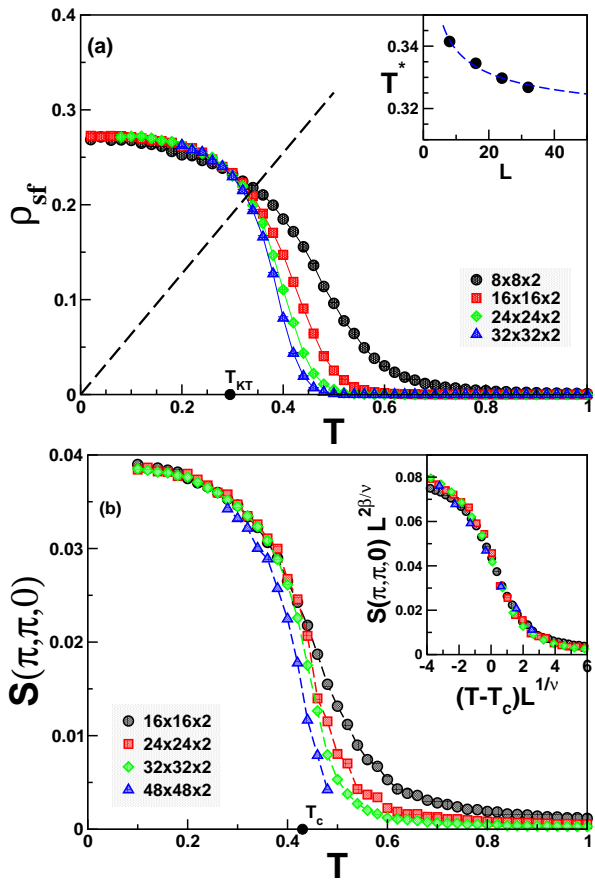

\begin{center}
\includegraphics[width=0.9\columnwidth,clip]{StiffnessSFH14}
\includegraphics[width=0.9\columnwidth,clip]{IsingH158}
\end{center}
\caption{(color online) Top panel (a): KT transition SF-paramagnet at $B=7$ [scan 3(a)]. Spin stiffness $\rho_{\rm sf}$ versus $T$ for various system sizes indicated by different symbols. The intersections with the dashed line at $2/\pi T^*$ are plotted in the inset versus $L$ and fitted to the logarithmic finite size scaling form (see text) with $T_{\rm{KT}}=0.295$ and $L_0=0.346$ (dashed blue curve). Bottom panel (b): Ising transition solid-paramagnet at $B=7.9$ [scan 3(b)]. Structure factor $S(\pi,\pi,0)$ versus $T$  for system sizes $16\le L\le 48$. The critical temperature $T_c\simeq 0.435$ is obtained by getting a data collapse for $s(\pi,\pi,0)L^{2\beta/\nu}$ as a function of $(T-T_c)L^{1/\nu}$ with $\beta=1/8$ and $\nu=1$, as shown in the inset.}
\label{KTIsing}
\end{figure}

At $T=0$ the SS phase is expected to display a true long range order for both diagonal and off-diagonal components, breaking respectively the lattice translation and the U(1) symmetries. On the other hand this is not true for $T>0$ since 
the continuous U(1) symmetry cannot be broken in 2D and only a quasi-long-range order is expected in the $xy$ plane. Nevertheless, the spin stiffness $\rho_{\rm sf}$ can remain finite up to a KT temperature $T_{\rm{KT}}$ and the diagonal order can persist at $T>0$ since it only breaks a discrete symmetry. Therefore, a persistence of the SS state at finite temperature is expected, but whether SF and solid components disappear simultaneously
or at different temperatures is not straightforward~\cite{XY}. As shown in Fig.~\ref{PhDgT}, we actually find that upon heating the melting of the SS occurs in two distinct steps: the SF density is destroyed via a KT transition while the solid order melts via an Ising transition at $T_c\neq T_{{\rm KT}}$. 

These critical properties of the SS are in fact intimately tied to the critical behavior of the surrounding SF and solid orders which are decoupled, and remain decoupled also in the SS regime.  
Indeed, as shown in Fig.~\ref{PhDgT}, the KT transition line between the SF and the paramagnet continues in the SS to eventually vanish at the SF-solid ($T=0$) quantum critical point. Similarly, the Ising transition line between the solid and the paramagnet is also prolongated down to the SF-SS quantum critical point.

The KT and 2D Ising natures of respectively the SF-paramagnet and solid-paramagnet transitions are illustrated in Fig.~\ref{KTIsing} where the scans 3(a) and 3(b) at $B=7$ and $B=7.9$ are shown. In Fig.~\ref{KTIsing} (a), the disappearance of the 2D SF density $\rho_{\rm sf}$ at $T_{\rm{KT}}$ is displayed for $L=8,16,24,32$. The KT transition, identified by a universal jump for an infinite system 
$
\rho_{\rm sf}(T_{\rm{KT}})=\frac{2}{\pi}T_{\rm{KT}}$,
suffers from logarithmic finite size corrections and the intersection occurs at 
$
T^*(L)=T_{\rm{KT}}[1+1/(2\ln(L/L_0))]
$,
as shown in the inset of Fig.~\ref{KTIsing} (a). 

\begin{figure}[!ht]
\begin{center}
\includegraphics[width=\columnwidth,clip]{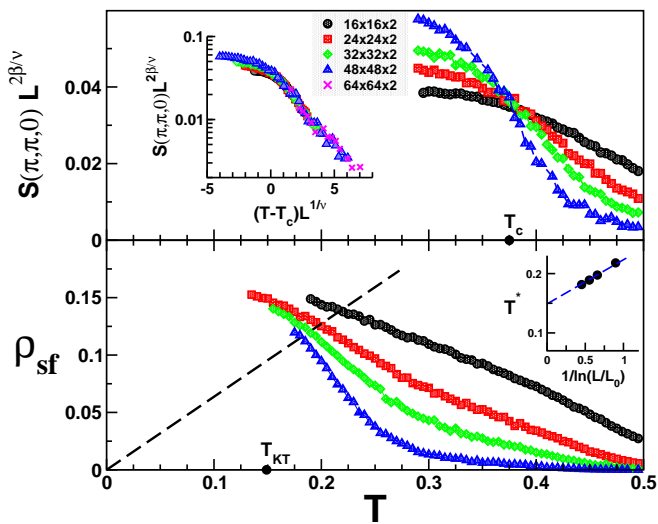}
\end{center}
\caption{(color online) Two different finite temperature transitions for the destruction of the SS phase at $B=7.3$ [scan 4]. Top panel: (2D Ising transition with $\beta=1/8$, $\nu=1$) structure factor plotted $S(\pi,\pi,0)L^{2\beta/\nu}$ versus $T$ for system sizes $16\le L\le 48$. Inset: Finite size scaling with $T_c=0.375$. Bottom panel: (KT transition) spin stiffness $\rho_{\rm sf}$ plotted versus $T$ for the same system sizes as in the top panel. The dashed line $2T/\pi$ intersects the data points for $T^{*}(L)$, plotted in the inset vs $1/\ln(L/L_{0})$ and fit to the logarithmic form (see text) with $T_{\rm{KT}}=0.149$ and $L_0=5.25$ (dashed blue line).}
\label{SST}
\end{figure}
Upon heating, the ($\pi,\pi,0$) solid melts via an Ising transition into a paramagnet at a critical temperature $T_c$, as shown in Fig.~\ref{KTIsing} (b). Using the finite size scaling hypothesis for the squared order parameter
$
S(\pi,\pi,0)=L^{-2\beta/\nu}{\cal{F}}\left[(T-T_c)L^{1/\nu}\right]$,
the Ising universality class is checked by getting a data collapse for $S(\pi,\pi,0)L^{2\beta/\nu}$ as a function of $(T-T_c)L^{1/\nu}$
for $L=8,\hdots,48$, with the Ising exponents $\beta=1/8$ and $\nu=1$ (inset of Fig.~\ref{KTIsing} (b)).
 
As already stated, the melting of the SS occurs in two steps, as illustrated in Fig.~\ref{SST} for $B=7.3$ (scan 4 of the phase diagram Fig.~\ref{PhDgT}). The universality classes (KT for the SF and Ising for the solid) are checked and the critical temperatures $T_{\rm KT}$ and $T_{c}$ are estimated. We notice that the logarithmic effects at the KT transition are more pronounced than for the SF to paramagnet transition since we found a length scale $L_{0}$ for the logarithmic correction much bigger than for the SF-paramagnet transition.
While upon heating the transitions SS $\to$ Solid $\to$ Paramagnet (scan 4 for instance) are clearly indentified as KT and Ising 2D, it is much harder to conclude on the nature of the finite $T$ transition  SS $\to$ SF. Indeed, even if 2D Ising is also expected, finite size effects turned out to be very large and it would require a huge computational effort to settle this issue.

\noindent
{\it{Conclusions---}}
We have confirmed the existence of the analog of a supersolid phase in a spin-1/2 bilayer, and 
we have fully characterized its temperature behaviour, which generically consists of two 
well separated phase transitions in the Kosterlitz-Thouless and Ising universality classes. 
We have also shown that a clear magnetization anomaly is expected upon entering the supersolid
phase from the superfluid one, and that the stiffness vanishes linearly at the edge of the plateau, in agreement with the superfluid-insulator universality class. 
These properties are expected to be relevant for other geometries, in 
particular
for systems where plateaus are produced by frustration rather then 
anisotropic
coupling, as recently agued in the context of a spin-1 model~\cite{Batista}. As such,
they should be useful in the search for supersolid phases in the 
neighborhood
of magnetization plateaus observed in some quantum antiferromagnetic
compounds like SrCu$_{2}$(BO$_{3}$)$_{2}$~\cite{Borate}. 

Transposed for bosons, the present results open a new route to supersolidity. Indeed,
to first order in $1/J_\perp$, the effective hard-core boson model is the simple $t-V$ model on the square
lattice, which is well known to have phase separation but no supersolid phase~\cite{Batrouni}. 
So the supersolid phase
has to emerge from higher terms, such as correlated hoppings, which were recently shown to induce a 
pairing phase at low density~\cite{Bendjama-Kay}.

Finally, from a more theoretical point of view, we have shown that the $T=0$ quantum phase transition from the superfluid to the supersolid is consistent with a 3D Ising universality class. 
However, to which extent our numerical data are consistent with the general expectation for a 
transition between a superfluid and a checkerboard supersolid~\cite{Balents} deserves further investigation.\\

We thank F. Alet, C. Berthier, A. Laeuchli, K. P. Schmidt, M. Takigawa and S. Wessel for interesting discussions. We also acknowledge the Swiss National Fund and MaNEP for financial support.


\begin{thebibliography}{99}
\bibitem{Penrose56-Andreev69} O. Penrose and L. Onsager, Phys. Rev. {\bf 104}, 576 (1956); A. F. Andreev and I. M. Lifshitz, JETP {\bf 29}, 1107 (1969).
\bibitem{Review} For a recent review, see N. Prokof'ev, Adv. Phys. {\bf 56}, 381 (2007).

\bibitem{Kim04} E. Kim and M. H. W. Chan, Nature (London) {\bf 427}, 225 (2004); Science {\bf 305}, 1941 (2004).

\bibitem{Balibar06} S. Sasaki, R. Ishiguro, F. Caupin, H. J. Maris and S. Balibar, Science {\bf 313}, 1098 (2006).


\bibitem{Pollet07} N. Prokof'ev {\it et al}., Phys. Rev. Lett. {\bf 94}, 155302 (2005) ; E. Burovskii {\it et al}., {\it ibid} {\bf 94}, 165301 (2005) ; M. Boninsegni {\it et al}., {\it ibid} {\bf 97}, 080401 (2006) ; L. Pollet {\it et al}., {\it ibid} {\bf 98}, 135301 (2007).

\bibitem{Triangular} S. Wessel and M. Troyer, Phys. Rev. Lett. {\bf 95}, 127205 (2005); D. Heidarian and K. Damle, {\it ibid.} {\bf 95}, 127206 (2005); R. Melko {\it {et al}}., {\it ibid.} {\bf 95}, 127207 (2005).
\bibitem{Prokof} M. Boninsegni and N. Prokof'ev Phys. Rev. Lett. {\bf 95}, 237204 (2005).
\bibitem{frederic98-giam99} F. Mila, Eur. Phys. J. B {\bf 6}, 201 (1998); T. Giamarchi and Tsvelik, Phys. Rev. B {\bf 59}, 11398 (1999).
\bibitem{hanpurple} M. Jaime {\it et al}.,
Phys. Rev. Lett. {\bf 93}, 087203 (2004); S. E. Sebastian {\it et al}., Nature {\bf 441}, 617 (2006).
\bibitem{TiOCl} M. Matsumoto {\it et al}., Phys. Rev. Lett. {\bf 89}, 077203 (2002); Phys. Rev. B {\bf 69}, 054423 (2004).
\bibitem{Oshikawa97} M. Oshikawa, M. Yamanaka, and I. Affleck
Phys. Rev. Lett. {\bf 78}, 1984 (1997). 
\bibitem{Borate} H. Kageyama {\it et al}.,
Phys. Rev. Lett. {\bf 82}, 3168 (1999); K. Kodama {\it et al}., Science {\bf  298}, 395 (2002).
\bibitem{Rice02} T. M. Rice, Science {\bf 298}, 760 (2002).
\bibitem{Senthil} T. Senthil {\it et al.}, Science {\bf 303}, 1490 (2004); T. Senthil {\it et al.}, Phys. Rev. B {\bf 70}, 144407 (2004). 
\bibitem{NgLee06} K.-K. Ng and T. K. Lee, Phys. Rev. Lett. {\bf 97}, 127204 (2006).
\bibitem{Sandvik02} O. Siljuasen and A. W. Sandvik, Phys. Rev. E {\bf 66}, 046701 (2002).
\bibitem{Ceperley-Kawashima} E. L. Pollock and D. M. Ceperley
Phys. Rev. B {\bf 36}, 8343 (1987); K. Harada and N. Kawashima
Phys. Rev. B {\bf 55}, 11949(R) (1997).
\bibitem{Blote95} H. W. J. Bl\"ote, E. Luijten  and J. R. Heringa, J. Phys. A: Math. Gen. {\bf 28}, 6289 (1995).
\bibitem{Fisher89-Alet04} M. P. A. Fisher {\it et al}., Phys. Rev. B {\bf 40}, 546 (1989); F. Alet and E. S. S\o rensen, Phys. Rev. B {\bf 70}, 024513 (2004).
\bibitem{XY} A similar issue has been addressed for the frustrated XY model in the context of Josephson-junction 
arrays, with the conclusion that both possibilities (a single transition or two transitions) can be realized depending on the microscopic details of the model. For a recent review, see S. E. Korshunov, Physics - Uspekhi {\bf 49}, 225 (2006).
\bibitem{Batista} P. Sengupta and C. D. Batista, Phys. Rev. Lett. {\bf 98}, 227201 (2007).
\bibitem{Batrouni} G. G. Batrouni and R. T. Scalettar, Phys. Rev. Lett. {\bf 84}, 1599 (2000).
\bibitem{Bendjama-Kay} R. Bendjama, B. Kumar, F. Mila, Phys. Rev. Lett {\bf 95}, 110406 (2005); K. P. Schmidt {\it et al}.,
Phys. Rev. B {\bf 74}, 174508 (2006).
\bibitem{Balents} E. Frey and L. Balents, Phys. Rev. B {\bf 55}, 1050 (1996).
\end{thebibliography}
\end{document}